\documentclass[12pt]{article}
\newcommand{\beq}{\begin{equation}}
\newcommand{\eeq}{\end{equation}} 

\begin{document}

\begin{center}
{\Large\bf Induced Lorentz and PCT Symmetry Breaking in an External
Electromagnetic Field} 
\vspace{1.0cm}
  
M. B. Hott and J. L. Tomazelli\footnote{e-mail: tomazell@dfq01.feg.unesp.br}
\\
\vspace{0.3cm}

{\em Departamento de F\'{\i}sica e Qu\'{\i}mica, Faculdade de Engenharia\\
Universidade Estadual Paulista, Campus de Guaratinguet\'a\\
Av. Dr. Ariberto Pereira da Cunha, 333\\
12500-000 Guaratinguet\'a, S. P., Brazil}
\vspace{0.7cm}
\end{center} 

\begin{abstract}
In this work we derive the Lorentz-PCT-violating effective action for a
fermion in a constant and uniform electromagnetic field using the
Fock-Schwinger proper time method and extract the exact value of the
coefficient of the nonperturbatively induced Chern-Simons term.
\end{abstract}  
\vspace{1.0cm}


Lattely, there has been increasing interest in extensions of quantum
electrodynamics (QED) and the Standard Model (SM), where Lorentz and PCT 
symmetries are broken in the fermion sector of the corresponding
Lagrangian density, both from phenomenological${}^{[1-3]}$ and 
field-theoretical \hfill \\ standpoints${}^{[4-6]}$.    

If we consider radiative corrections, the Lorentz- and PCT-violating axial 
vector term induces a four dimensional analogue of the so-called
Chern-Simons (CS) term in the free electromagnetic Lagrangian density
of the extended Maxwell electrodynamics, in much the same way as an
odd-parity mass term for fermions in QED in three-dimensional spacetime is
responsible for the topological mass term of the gauge boson${}^{[7]}$.

Nevertheless, in both cases, ambiguities occur in the usual perturbation 
theory, at leading order in the relevant expansion parameters: the
coefficient of the induced CS term is regularization dependent and
may only be fixed by physical requirements${}^{[8]}$. 

In this work we address the issue of generation of a Lorentz- and
PCT-violating CS term in the Lagrangian density of extended QED in (3+1)
dimensions, by considering a fermion in the presence of a constant and
uniform electromagnetic field. Employing the Fock-Schwinger proper time
construct${}^{[9]}$ together with the analytic regularization
method${}^{[10]}$ for the singular fermion Green's function, we
derive the renormalized effective action corresponding to the CS term and
extract its coefficient without any further approximations, confronting
our result with those obtained from nonperturbative calculations in
the sense of reference [6] and other recent approaches ${}^{[11,12]}$.      
             

The effective action for a fermion in the presence of an external
electromagnetic field in the extended version of QED considered in
reference [1] is defined by
\begin{eqnarray}
S_{\rm eff}(A) & = & -i \log  \int {\cal D} \overline{\psi} {\cal D} \psi 
\exp \left( i \int d^{4}x \, 
\overline{\psi} \ [ i \slash\!\!\!\partial -m - e A\!\!\!\slash(x) +
b\!\!\!\slash \gamma_{5} ] \ \psi \right)  \nonumber \\
& = & \log {\rm det} \left[ i \slash\!\!\!\partial -m - e A\!\!\!\slash(x)
+ b\!\!\!\slash \gamma_{5} \right] \ ,
\end{eqnarray}
\noindent where $b_{\mu}$ in the last term is a four-component {\em
constant} quantity which picks out a preferred direction in spacetime.
Such a term as it stands violates both Lorentz and PCT symmetry.

We are only interested in the CS sector, which is linear in $A_{\mu}(x)$
and arises from the induced vacuum polarization contribution to the total 
fermion current. This can be obtained according to the 
prescription${}^{[13]}$  
\beq
\left. \frac{\delta S_{\rm eff}}{\delta A_{\mu}(x)} = \langle j^{\mu}(x)
\rangle = ie \ {\rm tr} \left( \gamma_{\mu} \tilde{G}(x,x^{\prime})
\right) \right|_{x=x^{\prime}} \ , \label{current}
\eeq  
\noindent where $\tilde{G}(x,x^{\prime})$ is the gauge invariant part of
the fermion Green's function $G(x,x^{\prime})$, which satisfies the
inhomogeneous differential equation
\beq
\left[ i \partial\!\!\!\slash - m - e A\!\!\!\slash (x) + b\!\!\!\slash
\gamma_{5} \right] G(x,x^{\prime}) = \delta^{(4)}(x - x^{\prime} ) \ .
\label{eqmotion} 
\eeq


Now, we consider the Green's function 
\begin{eqnarray} 
G(x,x^{\prime}) & = & \langle x| [i \partial\!\!\!\slash - m - e A\!\!\!\slash
(x) + b\!\!\!\slash \gamma_{5}]^{-1} |x^{\prime} \rangle \nonumber \\
& = & \left(i \partial\!\!\!\slash + m - e A\!\!\!\slash
(x) + b\!\!\!\slash \gamma_{5} \right) S(x,x^{\prime}) \ ,
\end{eqnarray}

\noindent where $S(x,x^{\prime})$ in momentum space satisfies the equation
\beq
[i \pi\!\!\!\slash - m - e A\!\!\!\slash
(x) + b\!\!\!\slash \gamma_{5}][i \pi\!\!\!\slash + m - e A\!\!\!\slash
(x) + b\!\!\!\slash \gamma_{5}] \ S(p) = 1 \, 
\eeq

\noindent with $\pi_{\mu} = p_{\mu} - e A_{\mu}$. The above equation can
still be rewritten as
\beq
[\hat{\pi}^{2} - \frac{e}{2} \sigma .F - m^2 - 2b^{2}] \ S = 1 \ ,
\label{quadeq} 
\eeq

\noindent where $\hat{\pi}_{\mu} = \pi_{\mu} - i \sigma_{\mu \nu} b^{\nu}
\gamma_{5}$.

If we interpret the operator acting on $S$ in equation (\ref{quadeq}) as
the Hamiltonian of a quantum mechanical system, we can set the equations
of motion 
\begin{eqnarray}
\dot{x}_{\mu} &=& -i\left[ x_{\mu}, H \right] \,=\, -2\hat{\pi}_{\mu} \ ,  
\nonumber \\
\dot{\hat{\pi}}_{\mu} &=& -i\left[ \hat{\pi}_{\mu}, H \right] \,=\, -2e \
F_{\mu \nu}\hat{\pi}^{\nu} \ ,
\end{eqnarray} 

\noindent where the dotted operators mean derivatives with respect to a
proper time parameter $s$. This system admits the solution
\beq
\hat{\pi}_{\mu}(s)=-\frac{1}{2}\left[ \frac{e F {\rm e}^{-2eFs}}
{\sinh (eFs)} \right]_{\mu \nu} \ \Delta x^{\nu} \ ,
\eeq 

\noindent where $\Delta x^{\nu}=\left( x^{\nu}(s)-x^{\nu}(0) \right)$.
Hence, the Hamiltonian takes the form
\begin{eqnarray}
H &=& \Delta x^{\mu} K_{\mu \nu} \Delta x^{\nu} -2x^{\mu}(s) K_{\mu \nu}
x^{\nu}(0) -\frac{i}{2} {\rm tr}[e F \coth (eFs)] \nonumber \\
  &-& \frac{e}{2} \sigma .F - m^2 - 2b^2 \ ,
\end{eqnarray}

\noindent where $K_{\mu \nu}=\frac{1}{4}e^2F^2[\sinh^{-2}(eFs)]_{\mu \nu}$. 
Defining the evolution operator
\beq
U(x,x^{\prime};s)=\langle x|{\rm e}^{-iHs}|x^{\prime}\rangle \ ,
\eeq

\noindent we obtain
\begin{eqnarray}
U(x,x^{\prime};s) &=& \frac{C(x,x^{\prime})}{s^2} \exp \left\{ -\frac{1}{2} 
{\rm tr} \log \left[ (eFS)^{-1}\sinh (eFs) \right] \right\} \nonumber \\
&\times& \exp \left\{ \frac{i}{4} \Delta x \left[ eF\coth (eFs)  \right] 
\Delta x+\frac{ie}{2} \sigma .Fs+i(m^2-2b^2)s \right\} \nonumber \\
\nonumber \\
&=& \langle x(s)|x(0) \rangle \ ,
\end{eqnarray}

\noindent where $\Delta x_{\nu}=(x_{\nu}-x^{\prime}_{\nu})$.

From equation (\ref{quadeq}), we have in coordinate representation 
\beq
S(x,x^{\prime})=\langle x|H^{-1}|x^{\prime}\rangle \ .
\eeq
This Green's function is singular in the ultraviolet (UV) limit 
$x\rightarrow x^{\prime}$ and such singularity will manifest as an UV
logarithmic divergence in the Lorentz- and PCT-violating sector of the 
effective action. In order to perform calculations leading to renormalized
physical quantities, we adopt the analytic regularization scheme${}^{[10]}$,
with the replacement
\begin{eqnarray}
S(x,x^{\prime}) &\longrightarrow& S_{\lambda}(x,x^{\prime})=f_{\lambda}
\,m^{2\lambda}\langle x|\frac{1}{(H+i\epsilon)^{1+\lambda}}|x^{\prime}\rangle 
\nonumber \\
& = & \frac{f_{\lambda}\,m^{2\lambda}(-i)^{1+\lambda}}{\Gamma(1+\lambda)}
\langle x|\int_0^{\infty} ds\,s^{\lambda}e^{is(H+i\epsilon)}|x^{\prime}
\rangle \nonumber \\
& = & \frac{f_{\lambda}\,m^{2\lambda}(-i)^{1+\lambda}}{\Gamma(1+\lambda)}
\int_0^{\infty}ds\, s^{\lambda}\,U(x,x^{\prime};-s)e^{-s\epsilon} \ , 
\end{eqnarray} 

\noindent where the parameter $\lambda>-1$ ultimately goes to zero to recover
the original theory. In the above expressions $f_{\lambda}$ is an arbitrary
function of $\lambda$ which is equal to the unity for $\lambda = 0$. The
factor $m^{2\lambda}$ gives the correct dimension of the regularized 
propagator associated to the squared Hamiltonian $H$.       

We also note that
\beq
\left(i\partial_{\mu}-eA_{\mu}(x)\right)\langle x(-s)|x(0)\rangle = 
\langle x|\pi_{\mu}(-s)|x^{\prime}\rangle \ .
\eeq

\noindent Then, it follows from (8) that $C(x,x^{\prime})$ in (11) satisfies 
\beq
\left(i\partial_{\mu}-eA_{\mu}(x)-\frac{e}{2}F_{\mu \nu}(x-x^{\prime})^{\nu}
-i\sigma_{\mu \nu}\,b^{\nu}\gamma_5\right)C(x,x^{\prime})=0 \ ,
\eeq 

\noindent whose solution is 
\beq
C(x,x^{\prime})={\cal C}\Phi(x,x^{\prime}) \ ,
\eeq 
\noindent where we have defined
\beq
\Phi(x,x^{\prime}) \equiv \exp \left( -ie\int_{x^{\prime}}^x d\eta^{\mu}\,
\left[ A_{\mu}(\eta)+\frac{1}{2}F_{\mu \nu}(x-x^{\prime})^{\nu} \right] 
+\Delta x^{\mu}\sigma_{\mu \nu}\,b^{\nu}\gamma_5 \right) \ .
\eeq

\noindent From the normalization condition
\beq
\lim_{s\rightarrow 0} \langle x(s)|x^{\prime}(0)\rangle =
\lim_{s\rightarrow 0} \langle \tilde{x}(s)|\tilde{x}^{\prime}(0)\rangle =
\delta^{(4)}(x-x^{\prime}) \ , 
\eeq 
\noindent we obtain
\beq
{\cal C}=-\frac{i}{(4\pi)^2} \ ,
\eeq 
 
\noindent where $\tilde{x}_{\mu}=x_{\mu}+2is\sigma_{\mu \nu}b^{\nu}\gamma_5$. 

Thus, assuming that equation (4) holds for regularized Green's functions,
\beq
G_{\lambda}(x,x^{\prime})=e^{ie{\cal P}\int_{x^{\prime}}^{x}d\eta^{\mu} \, 
A_{\mu}(\eta)} \, \tilde{G}_{\lambda}(x,x^{\prime}) \ , 
\eeq

\noindent where
\begin{eqnarray}
\tilde{G}_{\lambda}(x,x^{\prime}) & = & \frac{-f_{\lambda}\,m^{2\lambda}
(-i)^{1+\lambda}}{\Gamma (1+\lambda)(4\pi)^2}\int_0^{\infty}ds\,
s^{\lambda -2} \, e^{-is(m^2 + b^2 - i\epsilon)} \nonumber \\
& \times & \left\{ \frac{1}{2}\Delta x^{\mu}\left( \left[ eF\coth(eFs) 
\right]_{\mu \nu} - eF_{\mu \nu} \right)\gamma^{\nu} - 2b\!\!\!\slash
\gamma_5 + m \right\} \nonumber \\
& \times & \frac{e\sqrt{2{\cal F}}s}{\sin(e\sqrt{2{\cal F}}s)} 
\left[\cos(e\sqrt{2{\cal F}}s) - 
\frac{i\sigma .\, F}{\sqrt{2{\cal F}}}\sin(e\sqrt{2{\cal F}}s) \right] 
\nonumber \\
& \times &  \exp \left\{ -\frac{i}{4s}
\Delta x_{\mu} \left[ eFs\coth(eFs) \right]^{\mu \nu}
\Delta x_{\nu} + \Delta x^{\mu}\sigma_{\mu \nu}b^{\nu}\gamma_5 \right\} \, .
\end{eqnarray}

\noindent with ${\cal F}=\frac{1}{4}F_{\mu \nu}F^{\mu \nu}$. This quantity 
appears together with 
$${\cal G}=\frac{1}{4}F_{\mu \nu}\tilde{F}^{\mu \nu}=\vec{E} .\,\vec{B}$$ 

\noindent if we expand the exponential factor of the product $\sigma .\,F$ in 
equation (11). We may choose $\vec{E} \bot \vec{B}$ such that the first 
exponential in (11) also simplifies, leading to the above result.    

Recalling equation (\ref{current}), the expectation value of
the regularized fermion current is found to be 
\begin{eqnarray}
\langle j_{\lambda}{}^{\mu}\rangle & = & ie \left. \ {\rm tr} 
\left( \gamma_{\mu} \tilde{G}_{\lambda}(x,x^{\prime}) \right) 
\right|_{x=x^{\prime}} \nonumber \\
& = & \frac{-ief_{\lambda}\,m^{2\lambda}(-i)^{1+\lambda}}{\Gamma (1+\lambda)
(4\pi)^2}\int_0^{\infty} ds\,s^{\lambda -2} \,e^{-is(m^2 + b^2 - i\epsilon)} 
{\rm tr}\left\{ (-2b_{\nu}\gamma^{\mu}\gamma^{\nu}\gamma_5) \right. 
\nonumber \\
&   & \left. \frac{e\sqrt{2{\cal F}}s}{\sin(e\sqrt{2{\cal F}}s)} \left[
\cos(e\sqrt{2{\cal F}}s) - \frac{i\sigma .\, F}{\sqrt{2{\cal F}}}
\sin(e\sqrt{2{\cal F}}s) \right] \right\} \nonumber \\
& = & \frac{2e^2f_{\lambda}\,m^{2\lambda}(-i)^{1+\lambda}}{\Gamma (1+\lambda)
(4\pi)^2}\int_0^{\infty} ds\,s^{\lambda -1}b_{\nu} \,{\rm tr}(\gamma_5
\gamma^{\mu}\gamma^{\nu}\sigma^{\alpha \beta})\,F_{\alpha \beta} 
\,e^{-is(m^2 + b^2 - i\epsilon)} \nonumber \\
& = & \frac{-ie^2f_{\lambda}\,m^{2\lambda}(-i)^{1+\lambda}}{\Gamma (1+\lambda)
 \pi^2}\int_0^{\infty} ds\,s^{\lambda-1} \,e^{-is(m^2 + b^2 - i\epsilon)}
 \,b_{\nu}\tilde{F}^{\nu \mu} \ .
\end{eqnarray} 

Therefore, the regularized CS effective action reads
\beq
S_{\lambda}^{CS}=a_{\lambda}\int d^4x \,b_{\nu}\tilde{F}^{\nu \mu}A_{\mu} \ ,
\eeq 

\noindent where
\begin{eqnarray}
a_{\lambda} & = & \frac{-ie^2f_{\lambda}\,m^{2\lambda}(-i)^{1+\lambda}}
{\Gamma (1+\lambda) \pi^2} \int_0^{\infty} ds\,s^{\lambda-1} 
\,e^{-is(m^2 + b^2 - i\epsilon)} \nonumber \\
& = & \frac{-ie^2f_{\lambda}\,m^{2\lambda}(-i)^{1+\lambda}}{\Gamma 
(1+\lambda) \pi^2}\frac{\Gamma (\lambda)}{i^{\lambda}
(m^2 + b^2 - i\epsilon)^{\lambda}} \ .
\end{eqnarray}

\noindent Expanding $a_{\lambda}$ in Laurent series around $\lambda = 0$ and
retaining the finite real part, we arrive at the renormalized CS effective
action, with coefficient
\beq
a=\frac{e^2}{\pi^2}\left[ \log \left(\frac{m^2}{m^2 + b^2}\right) 
+ f_0^{\prime} \right] \ ,
\eeq   
 
\noindent where $f_0^{\prime}$ stands for the derivative of $f_{\lambda}$ at  
$\lambda =0$. This completes our calculation. 


We have performed a nonperturbative calculation of the coefficient of the CS
contribution to the gauge-invariant effective action of a fermion in a
constant uniform electromagnetic field. From (25), we see that the CS
coefficient $a$ exhibit a logarithmic contribution, which is {\em analytic}
in $b^2$, but singular for $m\rightarrow 0$, in contrast with the opposite
behavior of the corresponding contribution found in reference [12], which is
non-analytic in $b^2$ and vanishes for $m\rightarrow 0$. The later situation
is expected neither in perturbation theory${}^{[4]}$, at leading order in 
$b_{\mu}$, nor in the lowest order approximation considered in [6] and [11]. 
 
In our case, for $b^2<<m^2$ we have
\[ a\sim\frac{e^2}{\pi^2}\left[ -\frac{b^2}{m^2} + f_0^{\prime} \right]\ .\]  

\noindent The first term in the last expression comes from the singular
sector of the CS effective action and is absent from the finite lowest 
order result derived in reference [6]. However, the gauge-invariant vacuum
polarization amplitude is logarithmically divergent and must be regularized 
{\em as a whole} object. This would lead to the above mentioned arbitrariness 
in the coefficient of the CS term. The same reasoning applies to the
calculation performed in reference [11].

The freedom of choice of $f_{\lambda}$ in the analytic regularization scheme
we have adopted is reflected by the presence of a residual contribution
proportional to $f_0^{\prime}$. This arbitrary constant is not determined
from the symmetries of the theory and can only be fixed through subsidiary
physical constraints set by experiment${}^{[1],[3]}$.

\section*{Acknowledgements}

One of the authors (JLT) thanks Conselho Nacional de Desenvolvimento
Cient\'{\i}fico e Tecnol\'ogico (CNPq, Brazil) for partial financial 
support.   
        	
\section*{References}

\begin{description}
\item[{[1]}] S. Carroll, G. Field and R. Jackiw, Phys. Rev. D {\bf 41},
1231 (1990).
\item[{[2]}] D. Colladay and V. A. Kosteleck\'y, Phys. Rev. D {\bf 58},
116002 (1998).  
\item[{[3]}] S. Carroll and G. Field, Phys. Rev. Lett. {\bf 79}, 2394
(1997). 
\item[{[4]}] J. -M. Chung and P. Oh, Phys. Rev D {\bf 60}, 067702 (1999).
\item[{[5]}] S. Coleman and S. Glashow, Phys. Rev. D {\bf 59}, 116008 (1999). 
\item[{[6]}] R. Jackiw and V. A. Kosteleck\'y, Phys. Rev. Lett. {\bf 82},
3572 (1999). 
\item[{[7]}] S. Deser, R. Jackiw and S. Templeton, Ann. Phys. (NY) {\bf
140} 372 (1982).
\item[{[8]}] B. M. Pimentel and J. L. Tomazelli, Prog. Theor. Phys. {\bf
95} 1217 (1996).
\item[{[9]}] V. Fock, Physik. Z. Sowjetunion {\bf 12}, 404 (1937); J. 
Schwinger, Phys. Rev. {\bf 82} 664 (1951). 
\item[{[10]}] C. G. Bollini, J. J. Giambiaggi and A. Gonzales Dominguez,
Nuovo Cimento {\bf 31}, 550 (1964); for further details, see also E. R.
Speer, {\em Renormalization Theory}, Erice 1975, G. Velo and A. S. Wightman
eds., D. Reidel, Dordrecht, Holland (1976).  
\item[{[11]}] J. -M. Chung, Phys. Rev. D {\bf 60} 127901 (1999).
\item[{[12]}] C. D. Fosco and J. C. Le Guillou, hep-th/9904138.
\item[{[13]}] A. Redlich, Phys. Rev. D {\bf 29}, 2366 (1984).
\end{description}

\end{document}